\documentclass[fleqn,usenatbib]{mnras}

\usepackage{newtxtext,newtxmath} 
\usepackage[T1]{fontenc} 
\usepackage{ae,aecompl}

\usepackage{graphicx}
\usepackage{amsmath,amssymb}

\usepackage[dvipsnames]{xcolor}
\title[Realistic gravitational focusing of meteoroids]{Realistic gravitational focusing of meteoroid streams}

\author[A.\ V.\ Moorhead et al.]{
Althea V.\ Moorhead,$^{1}$\thanks{E-mail: althea.moorhead@nasa.gov} Tiffany D.\ Clements,$^{2}$ and Denis Vida$^{3,4}$ \\
$^{1}$NASA Meteoroid Environment Office, Marshall Space Flight Center EV44, Huntsville, Alabama 35812, USA\\
$^{2}$Aerodyne Industries, Jacobs Space Exploration Group, Marshall Space Flight Center EV44, Huntsville, Alabama 35812, USA\\
$^{3}$Department of Earth Sciences, University of Western Ontario, London, Ontario, N6A 5B7, Canada\\
$^{4}$Department of Physics and Astronomy, University of Western Ontario, London, Ontario, N6A 3K7, Canada
}

\date{Accepted XXX. Received YYY; in original form ZZZ}

\pubyear{2020}

\begin{document}
\label{firstpage}
\pagerange{\pageref{firstpage}--\pageref{lastpage}}
\maketitle

\begin{abstract}
The number density and flux of a meteoroid stream is enhanced near a massive body due to the phenomenon known as gravitational focusing. The greatest enhancement occurs directly opposite the massive body from the stream radiant: as an observer approaches this location, the degree of focusing is unbound for a perfectly collimated stream. However, real meteoroid streams exhibit some dispersion in radiant and speed that will act to eliminate this singularity. In this paper, we derive an analytic approximation for this smoothing that can be used in meteoroid environment models and is based on real measurements of meteor shower radiant dispersion.
\end{abstract}

\begin{keywords}
meteors -- meteoroids
\end{keywords}

\section{Introduction}
\label{sec:intro}

As meteoroids approach a massive body such as the Earth, they are accelerated by the body's gravitational field. The initial trajectory is bent and the meteoroid accelerates. The number density near the massive body increases due to the inward deflection of meteoroids, and the flux increases further still due to the increase in speed. This phenomenon is known as gravitational focusing, and when trajectories are blocked by intersection with the massive body, it is called planetary shielding. Both effects must be taken into account in order to model the meteoroid environment near a massive or sizeable body.

A global flux enhancement factor can be derived by invoking conservation of energy and angular momentum \citep{1951PRIA...54..165O,1972NASTN6596.....K,2007MNRAS.375..925J}.  However, gravitational focusing is highly anisotropic \citep{Divine:1992uo,1997AdSpR..19..301S,2002dsso.conf..359M,2007MNRAS.375..925J} and the number density is at its highest directly opposite the meteoroid radiant. In fact, the anti-radiant line corresponds to a singularity in the gravitational focusing equations, in which one-dimensional rings of initial meteoroid trajectories are focused onto zero-dimensional points along the anti-radiant. As a result, any observer or spacecraft that wanders close to the anti-radiant is predicted to encounter a massively increased meteoroid flux. 

Real meteoroid streams will not be perfectly collimated; some dispersion in velocity will always be present. Numerical simulations have demonstrated that a radiant with some ``width'' results in finite gravitational focusing at all locations \citep{2007MNRAS.375..925J}. However, an analytical approximation for this effect is needed for efficient modeling of the meteoroid environment. In this paper, we derive a modified analytical treatment of gravitational focusing and planetary shielding that incorporates smoothing terms and we tie these modifications to observed dispersions in meteor shower radiants.  

There is a second singularity in the gravitational focusing equations: the flux or number density enhancement due to gravitational focusing tends to infinity as the meteoroids' initial speed relative to the massive body approaches zero. This behavior was noted by \cite{1951PRIA...54..165O} and is a result of the assumptions inherent in most treatments of gravitational focusing. These assumptions are that there are no other gravitational bodies present, that the stream of meteoroids is infinitely wide, and even that the universe is infinitely old. Under these conditions, a single massive body can indeed attract initially stationary meteoroids from arbitrarily large distances, resulting in an infinitely large flux. In this paper, we compute a velocity smoothing term that assumes gravitational focusing is limited by the massive body's Hill radius.


These two smoothing terms solve different meteoroid dynamics problems. Our angular smoothing term results in more realistic meteoroid fluxes near the anti-radiant; relevant scenarios include gravitational focusing of a meteor shower by the Earth at the Moon's location, by the Moon at the Earth's location, and by the Earth at the location of a high-orbiting satellite, such as one in geosynchronous orbit. Our velocity smoothing term results in more realistic fluxes at all angles for very slow meteoroids. This is more likely to be useful when modeling sporadic meteoroids or dust whose orbits have been circularized due to Poynting-Robertson drag and thus encounter the Earth at very low speeds.

One can, of course, directly simulate the motion of meteoroids in response to the Earth's gravity. This approach is accurate but computationally intensive. We thus presume that anyone applying these equations is doing so to reduce run time and so we prioritize numerical efficiency over fidelity. 


\section{Radiant dispersions}

A meteor shower originates from a single body, usually a comet. However, even if the parent body's orbit remains unchanged over time, meteoroids are propelled away from the parent by sublimating gases; this process alone may produce velocity dispersions of nearly 1~km~s$^{-1}$ in some cases \citep{2018M&PS...53.1292M}. Meteoroid orbits are further modified by planetary perturbations and by size-dependent radiation-related effects such as solar radiation pressure, Poynting-Robertson drag, and the Yarkovsky effect. The resulting dispersions in radiant and speed are then obscured by differential deceleration in the atmosphere and by measurement error \citep{1992CoSka..22..123K}. The true dispersion in speed is particularly difficult to measure, as precise measurements of the initial speed cannot be obtained without modeling the deceleration of each meteor in the atmosphere \citep{vida2018modelling}.

Meteor radiant measurements are less sensitive to atmospheric deceleration and can, if one assumes the velocity dispersion is isotropic, also be used to estimate the speed dispersion \citep{1970BAICz..21..153K,1998EP&S...50..555J}. Radiant dispersions are not routinely measured, but some estimates do exist in the literature.  \cite{1970BAICz..21..153K} presented meteor shower radiant dispersions for nine showers using high-quality double-station photographic data, and found that the observed radiant spreads were broadly consistent with a 1~km~s$^{-1}$ velocity dispersion. \cite{1997P&SS...45..853B} presented meteor radiants from the 1995 Leonid outburst that appear to exhibit a dispersion much less than 1$^\circ$. \cite{1998MNRAS.301..941J} compared Perseid radiant dispersions in typical and outburst years and found the latter was significantly tighter. Several older studies obtained very tight radiant dispersions (e.g., 6.2' for the 1946 Draconids) based on the intersections of single-station meteor observations \citep{1950ApJ...111..104J,1963mmc..book..674M}. We suspect this method, which averages many orbits, underestimates the true dispersion.

These historical measurements are sometimes limited by measurement error and, in many cases, do not reduce the observed dispersion to a single quantitative measurement. Therefore, in this section, we present real radiant dispersions for two showers observed using the Global Meteor Network and demonstrate that they can be described by a Rayleigh distribution. A more detailed analysis of these data will be the subject of a future paper and the distributions presented here are intended to be illustrative rather than definitive.

\subsection{Methods}

The Global Meteor Network\footnote{GMN website: \url{https://globalmeteornetwork.org/}} (GMN) is a video meteor network of stations using low-cost internet protocol (IP) cameras and Raspberry Pi single-board computers running open source software \citep{vida2019overview}. As of late 2019, there are more than 150 GMN cameras in 20 countries. The cameras use either Sony IMX291 or IMX307 CMOS sensors and are operated at 25 frames per second. Although the cameras support Full HD resolution (1920x1080 px), they are run at 1280x720 px to accommodate the computational and storage limitations of Raspberry Pi 3 B+ single-board computers. GMN stations mostly use either 3.6~mm or 16~mm lenses; see table~\ref{tab:lenses} for their properties and stellar limiting magnitudes. 
A network of 7 stations with 16~mm lenses is located in western Croatia and all observe the same volume of the sky to ensure high quality meteor trajectory measurements. We will use this 16~mm data as a reference high-quality data set and investigate whether systems with 3.6~mm lenses observe the same shower radiant dispersions.

\begin{table}
    \centering
    \begin{tabular}{lcc}
        focal length & 3.6~mm & 16~mm \\
        f-number & 0.95 & 1.0 \\
        field of view & $90^{\circ} \times 45^{\circ}$ &
        $20^{\circ} \times 10^{\circ}$ \\
        pixel size & 4'/px & 1'/px \\
        \hline
        limiting mag & $+6 \pm 0.5$ & $+9 \pm 0.5$ 
    \end{tabular}
    \caption{Characteristics and stellar limiting magnitudes of two common lens types used by GMN stations. The limiting magnitude varies due to sky conditions.}
    \label{tab:lenses}
\end{table}

All GMN stations run open source Raspberry Pi Meteor Station (RMS) software\footnote{RMS library on GitHub: \url{https://github.com/CroatianMeteorNetwork/RMS}}. The details of the meteor detection algorithm and the software in general are given in \cite{vida2016open, vida2018first}. The initial astrometric calibration is manually performed for all stations using the GAIA DR2 star catalog \citep{prusti2016gaia, brown2018gaia}. The camera pointing (field of view center, rotation, and scale) is automatically recalibrated on a block of 256 averaged video frames around every meteor detection. This is necessary due to a camera pointing drift of up to 10' per night from thermal expansion and contraction of the camera bracket. Note that the magnitude of the expansion is larger than the plate scale (up to ten times larger for the 16~mm systems), making recalibration an absolute necessity to obtain precise measurements.

The trajectories are computed using the Monte Carlo method described in \cite{vida2020mctheory}. Multi-station observations are paired based on temporal and spatial correlation, and only those solutions which have heights and velocities in the typical meteor range are taken. If a meteor is observed from three or more stations, observations from stations with angular fit residuals larger than $3\sigma$ from the mean are rejected and the trajectory is recomputed. From December 2018 until mid-November 2019, more than 35,000 orbits were collected using 77 GMN stations, and 2,500 orbits collected only using 16~mm systems. Trajectories computed using data from 3.6~mm systems have median trajectory fit residuals of about 60'', while trajectories from 16~mm system have about 10''. \cite{vida2019mcresults} show that in theory systems with 3.6~mm lenses should be able to achieve radiant measurement precision of about 0.1$^{\circ}$, which is much smaller than most modern multi-station estimates of radiant dispersions.

Meteors were associated with showers using the table of Sun-centered ecliptic shower radiant positions 
given in \cite{jenniskens2018survey}. This table, which is based on CAMS data, includes radiant drift and provides numerous, scattered sets of reference parameters per shower. For instance, within one degree of solar longitude near the Orionid peak, the table quotes 3894 reference radiants that span more than 15$^\circ$ in longitude and 11$^\circ$ in latitude. We assume that meteors within 1$^{\circ}$ in solar longitude, 3$^{\circ}$ in radiant, and 10\% in geocentric velocity of a shower reference location are members of that shower. If there are multiple shower candidates in the vicinity, the shower with the minimum closeness score $D_\mathrm{sce}$ is taken:
\begin{align}
    D_\mathrm{sce} &= \frac{| \Delta \lambda_{\odot}|}{1^\circ}+  \frac{\theta}{3^\circ} + \frac{|\Delta w|}{0.1 w}
\end{align}
\noindent where $\Delta \lambda_{\odot}$ is the difference in solar longitude between the meteor and the reference shower radiant, $\theta$ is the angular separation, and $\Delta w$ is the geocentric velocity ($w$) difference. 

Ideally, the 16~mm data would be used to estimate dispersions of all showers, but these stations produce a small fraction of the total number of meteors in the full data set. 
For instance, the 16~mm data set has only 51 Orionids, while the full data set has a total of 2384 Orionid meteors. To compensate, we apply aggressive quality cuts to the full data set: we keep only those meteors with (1) a convergence angle of at least $15^\circ$, (2) a median fit error no more than 60'', (3) an average velocity error no more than 2\%, and (4) a total angular radiant error no more than 0.25$^\circ$ for the Orionids or 0.35$^\circ$ for the Perseids. After these cuts, 632 Orionids and 566 Perseids remain in the full data set. In section \ref{sec:dispersion}, we use this filtered data set to characterize the radiant dispersion; we also compare this dispersion with that of the 16~mm data and show that the two data sets produce consistent results.

\subsection{Dispersion characterization}
\label{sec:dispersion}

\begin{figure}
    \centering
    \includegraphics[width=\linewidth]{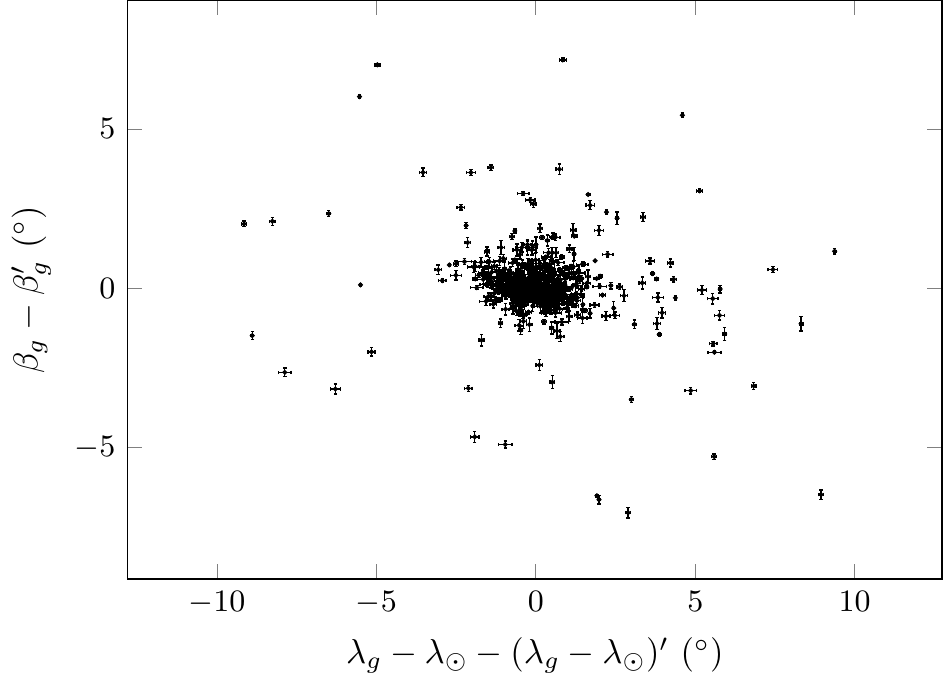}
    \caption{Position of 3.6~mm Orionid radiants relative to the average radiant (equations~\ref{eq:driftl} and \ref{eq:driftb}).}
    \label{fig:ori_spread}
\end{figure}

We present our geocentric meteor radiants in Sun-centered ecliptic longitude ($\lambda_g - \lambda_\odot$) and latitude ($\beta_g$) to minimize the movement of the shower radiant over time, but even in these coordinates some radiant drift is present. We must account for this drift in order to obtain the instantaneous radiant dispersion that will be relevant to our gravitational focusing calculations. We model the drift as a linear shift in both angular coordinates over time; for the Orionids, this drift is approximately:
\begin{align}
    (\lambda_g - \lambda_\odot)' &= 246.81^\circ - 0.25 \, (\lambda_\odot - 209^\circ) \label{eq:driftl} \\
    \beta_g' &= \, \, -7.64^\circ + 0.07 \, (\lambda_\odot - 209^\circ) \label{eq:driftb}
\end{align}
Subtracting equations~\ref{eq:driftl} and \ref{eq:driftb} from each Orionid meteor radiant results in a more compact distribution. Fig.~\ref{fig:ori_spread} presents the position of these radiants relative to the mean and includes their uncertainties as error bars. Note that the uncertainty in any individual meteor radiant is much smaller than the radiant dispersion, and thus measurement uncertainty does not make a significant contribution to the apparent dispersion.


\begin{figure}
    \centering
    \includegraphics[width=\linewidth]{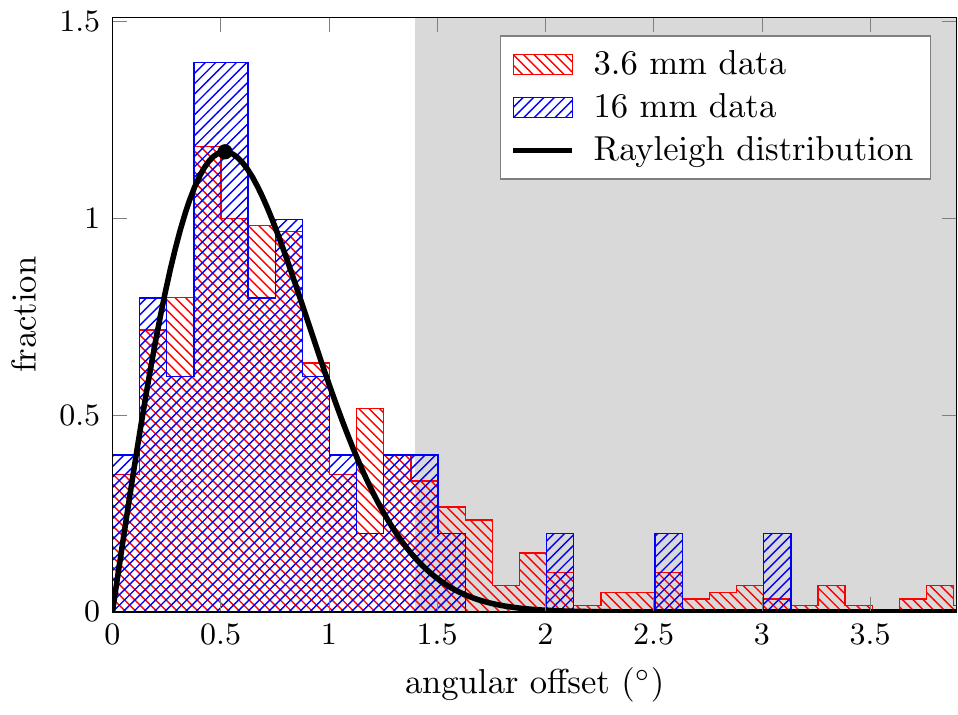}
    \caption{Angular offset of Orionid meteors relative to the mean radiant (see equations~\ref{eq:driftl} and \ref{eq:driftb}). Meteors more than 1.4$^\circ$ from the mean radiant (gray region) are excluded from the Rayleigh distribution fit. The mode of 0.52 is indicated by a black dot.}
    \label{fig:ori_hist}
\end{figure}

We next computed the angular offset of each radiant from the average; see Fig.~\ref{fig:ori_hist}. We found that the entire distribution was not well-described by a Rayleigh distribution because of the long tail of radiant offsets larger than $1-2^\circ$. We expect that this tail is due to contamination by sporadics or nearby showers. We eliminated these contaminants in a crude fashion by simply discarding those meteors that lay outside a cutoff value; we found that when we used a cutoff value of 1.4$^\circ$, the remaining Orionid meteors agreed well with a Rayleigh distribution. We therefore iterated our radiant drift fit to exclude meteors lying outside this boundary (equations~\ref{eq:driftl} and \ref{eq:driftb} include this refinement). We also restricted our data to include only those meteors falling within 15$^\circ$ of the median solar longitude for the shower.

In Fig.~\ref{fig:ori_hist}, we fit a Rayleigh distribution with a mode of 0.52$^\circ$ to our 3.6~mm Orionid data. A Kolmogorov-Smirnov (K-S) test fails to reject the hypothesis that the data follow a Rayleigh distribution with a p-value of 0.6. A second K-S test fails to reject the hypothesis that the 16~mm data follow this same distribution (with a p-value of 0.5). 
We therefore conclude that a Rayleigh distribution is a reasonable description of these data, and that the 3.6~mm and 16~mm data exhibit the same Orionid radiant dispersion. These values are comparable to those of \cite{1970BAICz..21..153K}, who give a median dispersion of $0.71^\circ$ for the Orionids and $1.07^\circ$ for the Perseids.

\begin{figure}
    \centering
    \includegraphics[width=\linewidth]{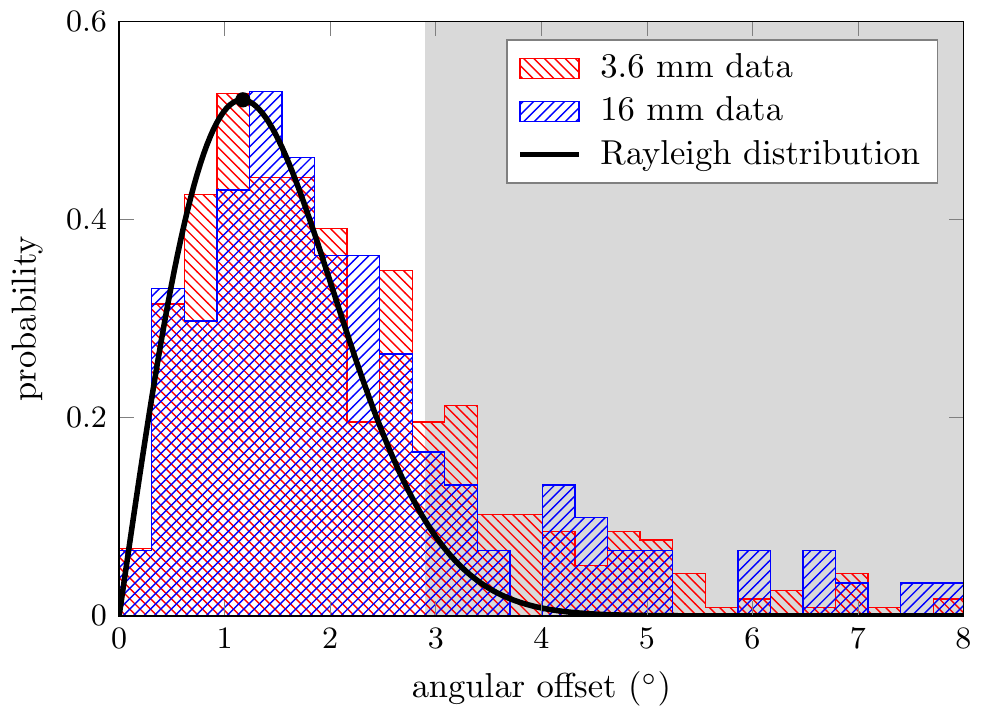}
    \caption{Angular offset of Perseid meteors relative to the mean radiant. Meteors more than 2.9$^\circ$ from the mean radiant (gray region) are excluded from the Rayleigh distribution fit. The mode of 1.2 is indicated by a black dot.}
    \label{fig:per_hist}
\end{figure}

We applied a similar set of calculations to a set of Perseid meteors. The Perseids exhibit less of a radiant drift, with a slope of only -0.03 in longitude and 0.14 in latitude. They also display a higher dispersion in their radiant (see Fig.~\ref{fig:per_hist}), and we obtained the closest resemblance to a Rayleigh distribution in this case by applying a radiant offset cutoff of 2.9$^\circ$. Here, the p-value for our Rayleigh distribution test between the 3.6~data and our best fit was 0.3, and the p-value between the 16~mm data and our best fit was 0.6. The mode of our best fit Rayleigh distribution was 1.2$^\circ$.

In this section we have demonstrated that, for both the Perseids and Orionids, the core component of the radiant dispersion can be described using a Rayleigh distribution (see appendix \ref{apx:dists} for additional discussion of the speed and radiant distributions). In both cases, the ``location'' of the Rayleigh distribution is fixed to 0, and so the only free parameter is the mode or scale of the distribution. Thus, we are now able to describe a meteor shower's radiant distribution with a single value.
A more careful subtraction of the sporadic contamination from these radiant dispersions is slated for a future paper. In the meantime, we conclude that our simulated Rayleigh distribution with a mode of just over 1$^\circ$ is a reasonable illustrative case. 

\section{Numerical simulations}
\label{sec:sims}

Now that we have quantified several shower radiant distributions, we next investigate the impact of such a distribution on gravitational focusing. We do so by conducting a series of numerical simulations of meteoroid trajectories near a massive body. In these simulations, meteoroids approach a massive body on parallel or nearly parallel initial trajectories, and we sample their position at random times in order to determine the number density of meteoroids as a function of position.

\begin{figure}
    \centering
    \includegraphics{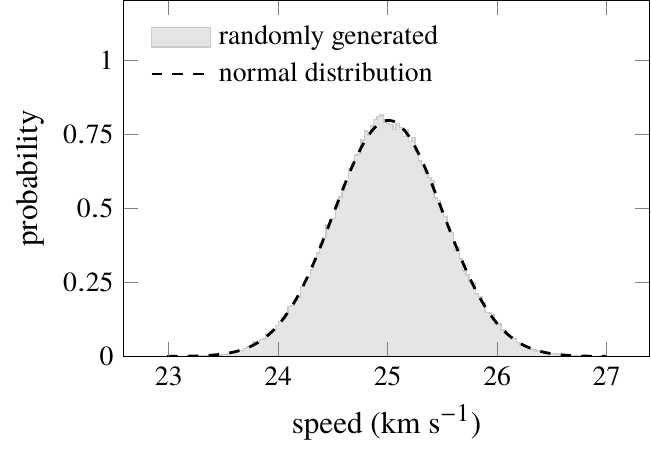}
    \caption{Simulated speed distribution (light gray histogram) corresponding to meteoroid velocity vectors in which each Cartesian component is described by a normal distribution with a standard deviation of 0.5~km~s$^{-1}$ and a mean of 25~km~s$^{-1}$. The distribution of the magnitude of the velocity is also well described by a normal distribution with a standard deviation of $\sigma = 0.5$~km~s$^{-1}$ (dashed black line).}
    \label{fig:vhist}
\end{figure}

\begin{figure}
    \centering
    \includegraphics{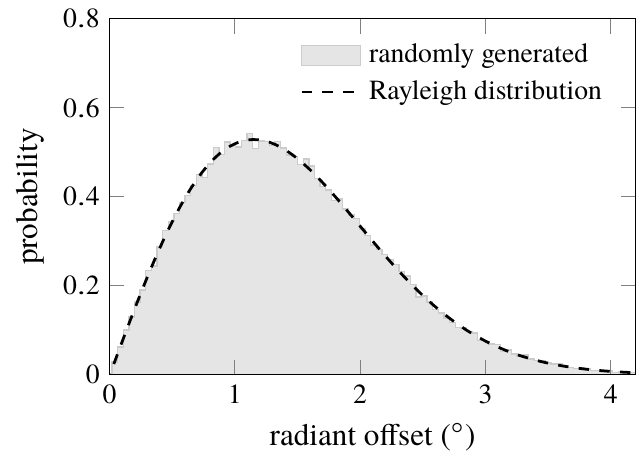}
    \caption{Simulated radiant dispersion (light gray histogram) corresponding to meteoroid velocity vectors in which each Cartesian component is described by a normal distribution with a standard deviation of 0.5~km~s$^{-1}$ and a mean of 25~km~s$^{-1}$. The dispersion is well described by a Rayleigh distribution with a mode of $\sigma = 1.15^\circ$ (dashed black line).}
    \label{fig:radhist}
\end{figure}

We construct four scenarios to explore the effects of meteoroid speed and radiant dispersions on gravitational focusing and planetary shielding. In the first scenario, no dispersions are present: meteoroids are placed on perfectly parallel trajectories far from the massive body. In the last scenario, we include an idealized meteoroid velocity dispersion in which the distribution of each component of the meteoroids' velocity vector is described by a Gaussian with a standard deviation of 0.5~km~s$^{-1}$. We construct two intermediate scenarios by dividing this velocity dispersion into two components: a speed dispersion and a radiant dispersion (see Fig.~\ref{fig:vhist} and \ref{fig:radhist}). In the former, we preserve only the variation in meteoroid speed and enforce a uniform direction of movement. In the latter case, we preserve only the variation in meteoroid directionality and enforce a constant speed. This approach assumes that the dispersion in meteoroid velocity is isotropic, and that the dispersion in speed can be determined from the dispersion in radiant and vice versa (see appendix~\ref{apx:dists}).

In all cases, we convert the initial trajectories of the meteoroids to orbital elements using the methodology of \cite{1999ssd..book.....M}, solve Kepler's equation \citep{1988CeMec..44..267G} at random times, and convert the orbital elements back to positions at those times. We then bin these positions in $x$, $y$, and $z$ and plot the $z=0$ slice of the grid. 

\begin{figure}
    \centering
    \includegraphics[width=\linewidth]{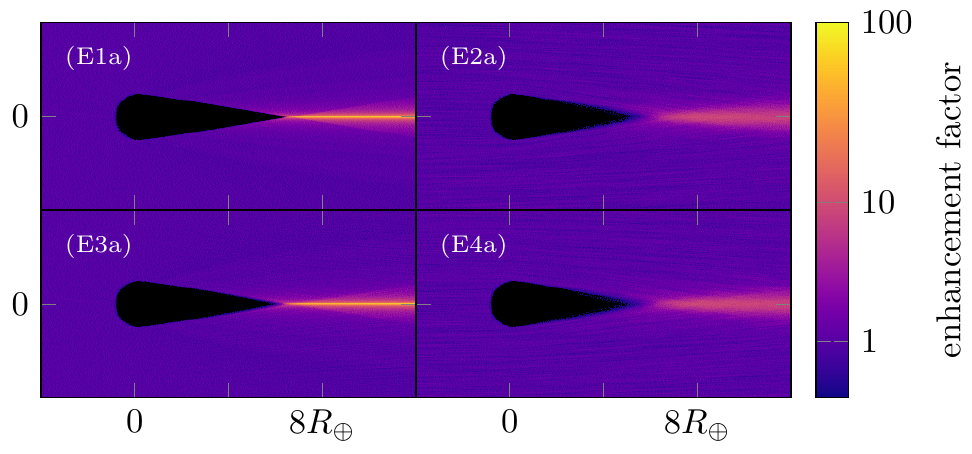}
    \caption{Number density of meteoroids near a massive body. In these simulations, meteoroids flow toward the massive body from the left and in the plane of the plot in (E1a) parallel, (E2a) with a radiant dispersion, (E3a) with a speed dispersion, and (E4a) with both a radiant and speed dispersion. The color scale reflects the logarithm of the relative number density of meteoroids; black regions contain no meteoroids.
    }
    \label{fig:comparison}
\end{figure}

A set of simulations is shown in Fig.~\ref{fig:comparison}. In these simulations, the meteoroids have an initial speed of 25~km~s$^{-1}$; we have selected this speed, which is significantly slower than the geocentric speed of the Orionids or Perseids, in order to obtain a more compact illustration of gravitational focusing. The massive body is assumed to be the Earth and we include 100~km of meteoroid-blocking atmosphere in the Earth's shielding radius. The color scale provides the number density enhancement factor, or ratio of the local meteoroid number density to that at infinity.  The massive body itself and its ``shadow'' -- those locations that are shielded by the massive body from the meteoroid stream -- appear in black.

When the meteoroid stream has no initial velocity or radiant dispersion (panel E1a of Fig.~\ref{fig:comparison}), the number density can be enhanced by two orders of magnitude near the anti-radiant line. The maximum value in our scale is limited only by our bin resolution (which is, in these simulations, 20 bins per $R_\oplus$); a finer resolution would produce even higher enhancement values. Introducing a speed dispersion only (panel E3a) causes the ``shadow'' of the massive body to end less abruptly but does not eliminate the anti-radiant singularity. Introducing a radiant dispersion (panel E2a), however, reduces the maximum enhancement factor to a factor of a few. We find that introducing a radiant dispersion only produces a result that is quite similar to that obtained with our full three-dimensional velocity dispersion (panel E4a). From these simulations, we conclude that, for the purpose of calculating gravitational focusing, it is more useful to measure a meteor shower's radiant dispersion than it is to measure its velocity dispersion.

\begin{figure*}
    \centering
    \includegraphics[width=6.67in]{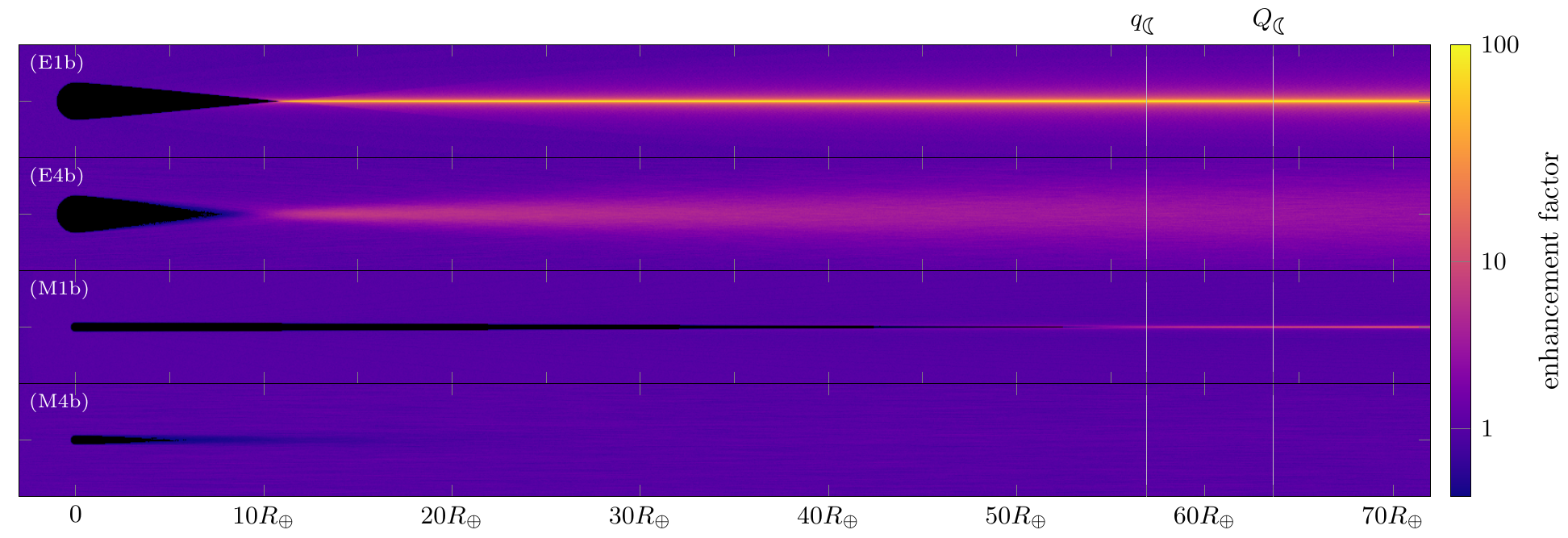}
    \caption{Number density of meteoroids near a massive body. In this case, we have simulated the gravitationally focused flow of meteoroids over a much longer distance; the blue dotted vertical lines mark the apogee and perigee distance of the Moon.}
    \label{fig:gem}
\end{figure*}

We present a second set of simulations in Fig.~\ref{fig:gem}. In this case, we set the initial speed of the meteoroid stream to 35~km~s$^{-1}$ and compute the number density over a distance that exceeds lunar perigee. We also include simulations in which we have used the Moon, rather than the Earth, as the massive body (panels M1b and M4b). For each massive body, we simulate the gravitational focusing of meteoroids on perfectly parallel paths (panels E1b and M1b) and those with a velocity dispersion of 0.5~km~s$^{-1}$ (and thus both a radiant and speed dispersion; panels E4b and M4b).

This simulation is designed to mimic the Geminid meteoroid stream, which has a geocentric speed of 35~km~s$^{-1}$ and lies near the ecliptic plane. Because of these properties, it has been suggested that the Moon could cause an enhancement of the Geminid flux visible at Earth via gravitational focusing (M.\ Matney, personal communication). Fig.~\ref{fig:gem}(M1b) does indeed suggest that the Moon could produce a narrow but significant enhancement of the Geminid flux near the Earth.
Based on this, one might expect short-lived, intense bursts of meteor shower activity whenever the Moon passes near the shower radiant. However, Fig.~\ref{fig:gem}(M1b) indicates that the Moon and Earth would have to be very closely aligned with the Geminid radiant, while in fact the Moon never passes within five degrees of the radiant. Furthermore, Fig.~\ref{fig:gem}(M4b) shows that a modest velocity dispersion of 0.5~km~s$^{-1}$ (which corresponds to a radiant dispersion of 0.8$^\circ$ at 35~km~s$^{-1}$) obliterates any enhancement, and also substantially shortens the ``shadow.'' Based on these results, we conclude that it is unlikely that the Moon could produce detectable flux enhancements in any meteor shower seen at Earth. However, Fig.~\ref{fig:gem}(E4b) shows that the Earth could produce noticeable enhancements at the Moon's location if the geometry is favorable.

The simulations shown in Fig.~\ref{fig:gem} also demonstrate that for less massive bodies, we will need to consider the effect of radiant dispersion on planetary shielding as well as on gravitational focusing. A modest radiant dispersion can drastically reduce the maximum distance at which a spacecraft can employ shielding as a sheltering technique, for instance.

\section{Analytical approximations}

The effects of gravitational focusing and shielding on a parallel stream of meteoroids can be computed analytically \citep{Divine:1992uo,1997AdSpR..19..301S,2002dsso.conf..359M,2007MNRAS.375..925J}. This approach is far more computationally efficient than simulating a large number of individual meteoroid orbits. As a result, those meteoroid environment models that describe the flux near planets employ an analytical approach \citep{Smith:1994wa,1997AdSpR..19..301S,Moorhead2019b}. In this section, we concoct modified versions of the \cite{1997AdSpR..19..301S} equations that incorporate the effect of radiant and velocity dispersion while preserving numerical efficiency. The equations for computing gravitational focusing and shielding are scattered throughout this section; for the convenience of the reader, they are also presented in a more compact form in appendix~\ref{apx:alg}.


\subsection{Gravitational focusing}

If the velocity of a meteoroid relative to a massive body before entering its gravitational well is $\vec{w}$, then the \emph{vis viva} equation dictates that its speed at a position $\vec{r}$ relative to the center of the body is
\begin{align}
    v &= \sqrt{w^2 + \frac{2 G M}{r}}
    \label{eq:veq}
\end{align}
where $G$ is the gravitational constant and $M$ is the mass of the massive body. If we introduce the dimensionless variable $f = G M / r w^2$, equation~(\ref{eq:veq}) can be re-written as:
\begin{align}
    \frac{v^2}{w^2} &= 1 + 2 f \, .
\end{align}
Note that our $f$ is the inverse of the variable $F$ used by \cite{2007MNRAS.375..925J}.

\begin{figure}
    \centering
    \includegraphics[width=\linewidth]{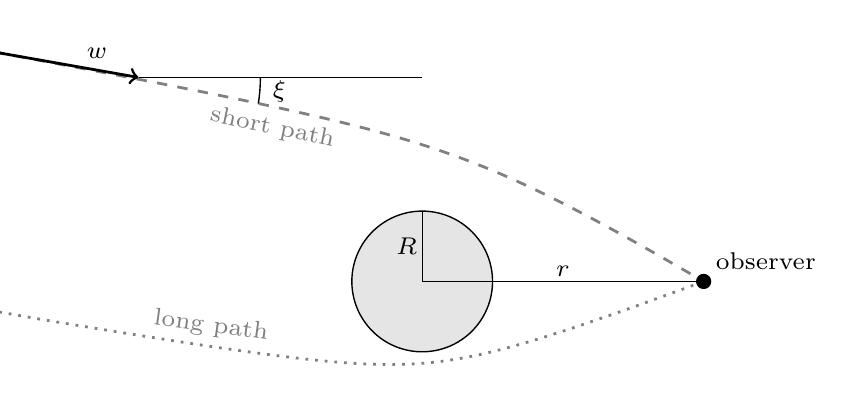}
    \caption{Diagram of a massive body (gray circle), observer location, and the two paths by which meteoroids from a given radiant and with initial speed $w$ can reach the observer.}
    \label{fig:diagram}
\end{figure}

Let $\xi$ represent the observer's angular offset from the anti-radiant; then, $w_r = w \cos \xi$ is the component of the un-focused meteoroid velocity parallel to the line between the massive body and the observer. Using these parameters, we can re-write equation~7 of \cite{1997AdSpR..19..301S} to obtain the enhancement of the meteoroid number density due to gravitational focusing, $\eta$:
\begin{align}
    \eta &= \frac{1}{2} \left| \, 1 \pm 
        \left. \left( 
            {\sin^2 \tfrac{\xi}{2} + f} 
        \right) \middle / b \right. \, \right| \, \textrm{, where} 
        \label{eq:etag} \\
    b &= \sqrt{
            \sin^2 \tfrac{\xi}{2}
            \left(\sin^2 \tfrac{\xi}{2} + 2 f \right)
        } \label{eq:b}
\end{align}
Note that our $b$ is equivalent to the unitless value $B$ of \cite{2007MNRAS.375..925J}, which, if multiplied by $w$, is the ``auxiliary velocity'' also named $B$ by \cite{1997AdSpR..19..301S}.
The ``$\pm$'' sign is present in $\eta$ because there are two paths by which meteoroids with a given radiant can reach a given point in space: one long, one short (see Fig.~\ref{fig:diagram}). The upper choice of sign corresponds to the short path, while the lower choice corresponds to the long path. Both must be included to determine the total enhancement. 

\subsubsection{Removal of the anti-radiant singularity}

Putting the equation for gravitational focusing in this form reveals its singularities and opportunities for smoothing over those singularities. For instance, $\eta$ tends toward infinity as $\xi \rightarrow 0$; this results in an infinitely large enhancement along the anti-radiant, as discussed in \S~\ref{sec:intro}. One can remove this singularity by introducing a minimum angle $\xi_\mathrm{min}$: 
\begin{align}
    \eta'_F &= \frac{1}{2} \left| \, 1 \pm 
        \left. \left( 
            {\sin^2 \tfrac{\xi}{2} +
            \sin^2 \tfrac{\xi_\mathrm{min}}{2}+ f} 
        \right) \middle / b' \right. \, \right| \, \textrm{, where} 
        \label{eq:etagp} \\
    b' &= \sqrt{
        \left( \sin^2 \tfrac{\xi}{2}
            + \sin^2 \tfrac{\xi_\mathrm{min}}{2}
        \right)
        \left(\sin^2 \tfrac{\xi}{2} +
        \sin^2 \tfrac{\xi_\mathrm{min}}{2} + 2 f \right)
        } \label{eq:bp}
\end{align}
This limits the enhancement near the anti-radiant while ensuring that $\eta'_F \simeq \eta$ when $\xi$ is large. Including $\sin^2 \xi_\mathrm{min}/2$ in the numerator of $\eta'$ as well as the denominator ensures that the total focusing factor tends to one when f and $\xi$ are both small, replicating the behavior seen in Fig.~\ref{fig:gem}(M4b).

\subsubsection{Removal of the slow meteoroid singularity}

The modifications made in equations~(\ref{eq:etagp}) and (\ref{eq:bp}) remove the anti-radiant singularity in the focusing factor, but a second singularity remains.  As the speed $w$ approaches zero, both $f$ and the focusing factor $eta$ approach infinity.  This corresponds to an idealized scenario in which the massive body is the only gravitational body in existence. In such a case, the massive body is capable of attracting infinitely distant meteoroids so long as they have no initial velocity with respect to the planet. This is unrealistic for several reasons. First, meteoroid streams are not infinitely broad. Second, the gravitational influence of planets in a solar system is limited to their Hill sphere. In the case of the Earth, the Hill sphere is generally smaller than a meteoroid stream 
\citep[although there can be exceptions; for instance][predict Draconid filaments narrower than 0.01~au]{2019Icar..330..123E}.
The Earth's Hill radius corresponds to a ``smoothing speed'' of 48~m~s$^{-1}$. This is much smaller than the speed of any meteor shower, but could be used to place an upper limit on the focusing of a highly circularized population of sporadic meteoroids or dust particles.

\subsection{Planetary shielding}
\label{sec:shielding}

Equation~(\ref{eq:etag}) or (\ref{eq:etagp}) describe how a massive body's gravity modifies the local meteoroid number density, but, for many geometries, the massive body may block one or both trajectories by which a meteoroid could otherwise reach the observer. This is termed ``planetary shielding'' and must also be taken into account. 

A trajectory is blocked if the meteoroid encounters the observer post-periapse \emph{and} the periapse distance lies inside the massive body's meteoroid-blocking radius.
\cite{1997AdSpR..19..301S} provide equations for computing the periapse distance and true anomaly of the meteoroid at the observer's location that can be used to test these conditions. We provide simplified versions of these equations in our unitless terms.

First, rather than calculate the true anomaly, which requires us to perform several numerically expensive square roots and an arcsine, we instead calculate the radial component of the meteoroid's speed at the observer's location and determine its sign:
\begin{align}
    \frac{v_r}{w} &= u_r = \cos^2 \tfrac{\xi}{2} \mp b 
        \label{eq:ur} > 0
\end{align}
where the upper choice of sign again corresponds to the short path. If equation~(\ref{eq:ur}) is true, then the meteoroid is moving away from the massive body at the observer's location and has thus passed through periapse.

Second, the periapse distance $q$ lies within the massive body's radius $R$ when 
\begin{align}
    \left( \frac{v_\theta}{w} \right)^2 &=
    u_\theta^2 = 1 + 2 f - \left( \frac{v_r}{w} \right)^2 
    < \left( \frac{R}{r} \right)^2 + 2 f \frac{R}{r}
    \label{eq:peri}
\end{align}
If both equations~(\ref{eq:ur}) and (\ref{eq:peri}) are satisfied, the path is blocked and makes no contribution to the local number density.

\begin{figure}
    \centering
    \includegraphics[width=\linewidth]{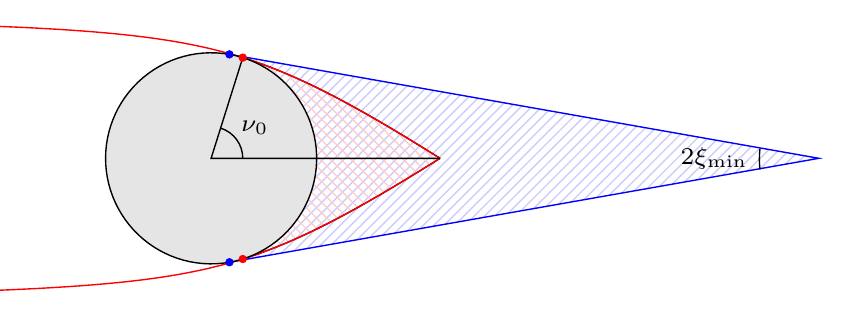}
    \caption{Diagram of the nominally shielded region (red) and the ``dispersion triangle'' (blue). Note that the lines in red are curved and thus the nominally shielded region is not a triangle. In this case, the shielded region lies within the dispersion triangle.}
    \label{fig:triangle}
\end{figure}

We were unable to concoct a simple modification of these equations that mimics the blurring effect that a radiant dispersion has on the planetary shielding pattern. In the case of the Earth, such a modification is probably unnecessary, as the shape of the ``shadow'' is very similar in the presence or absence of a radiant dispersion. In cases where planetary shielding plays a more prominent role than gravitational focusing, however, some alternative treatment of the shadow is clearly needed (see, e.g., panel M4b of Fig.~\ref{fig:gem}). In this case, we use the following shielding requirement:
\begin{align}
    \sin \xi + \cos \xi \, \tan \xi_\mathrm{min} &< \frac{R}{r} \label{eq:ylim}
\end{align}
When this equation is satisfied, the observer lies within an isosceles triangle whose two equal sides are tangent to the massive body and where the angle between them is $2 \xi_\mathrm{min}$ (see Fig.~\ref{fig:triangle}). We assume that the ``shadow'' can be no larger than this ``dispersion triangle.'' 

The repeated evaluation of this additional equation may be undesirable, especially for cases such as that depicted in Fig.~\ref{fig:gem}(M4b), where the shielded region is clearly smaller than the triangle defined by $\xi_\mathrm{min}$. In such a case, it may be numerically advantageous to perform one calculation at the outset to determine whether (a) the nominally shielded region has a smaller radial extent than the dispersion triangle, or (b) the nominally shielded region lies entirely within the dispersion triangle.

\subsubsection{The shielding-dominated case}

Let us use $r_0$ to denote the largest distance from the center of the massive body at which the observer can experience classical planetary shielding. This distance is given by:
\begin{align}
    r_0 &= R - \frac{R^2}{2 a}
\end{align}
where $a = -G M/w^2$ is the semi-major axis of the meteoroids. Condition (a) is therefore equivalent to:
\begin{align}
    \sin \xi_\mathrm{min} &> \frac{2 a}{2 a - R}
    \label{eq:conda}
\end{align}
When the above inequality is true, the dispersion triangle lies entirely within the nominal shadowing region. If equation~(\ref{eq:ylim}) is satisfied, we assume the short path is blocked. Contributions from the long path can be neglected, as $\eta' \simeq 0$ when the lower choice of sign is taken and $f$ is small compared to $\sin^2 \xi_\mathrm{min}/2$.

\subsubsection{The focusing-dominated case}

The true anomaly of a meteoroid that grazes the massive body and intersects the observer along the anti-radiant line is given by $\cos \nu_0 = a/(a-R)$. Thus, the nominally shadowed region lies entirely within the dispersion triangle when
\begin{align}
    \sin \xi_\mathrm{min} &< \frac{a}{a-R} \label{eq:gravdom}
\end{align}
In this case, gravitational focusing determines the shape of the shadowed region and it is unnecessary to evaluate equation~(\ref{eq:ylim}).

\subsubsection{Intermediate cases}
\label{sec:intermediate}

Because meteoroids passing near the massive body do not follow straight paths, cases exist in which neither the nominally shielded region nor the dispersion triangle lie entirely within the other. This occurs when:
\begin{align}
    \frac{a}{a-R} &< \sin \xi_\mathrm{min} <
    \frac{2 a}{2 a - R}
\end{align}
In this case, one must evaluate both equations~(\ref{eq:ur}) and (\ref{eq:ylim}), as well as equation~(\ref{eq:peri}), to determine whether the observer lies within the shielded region. 

\begin{figure*}
    \centering
    \includegraphics[width=6.92in]{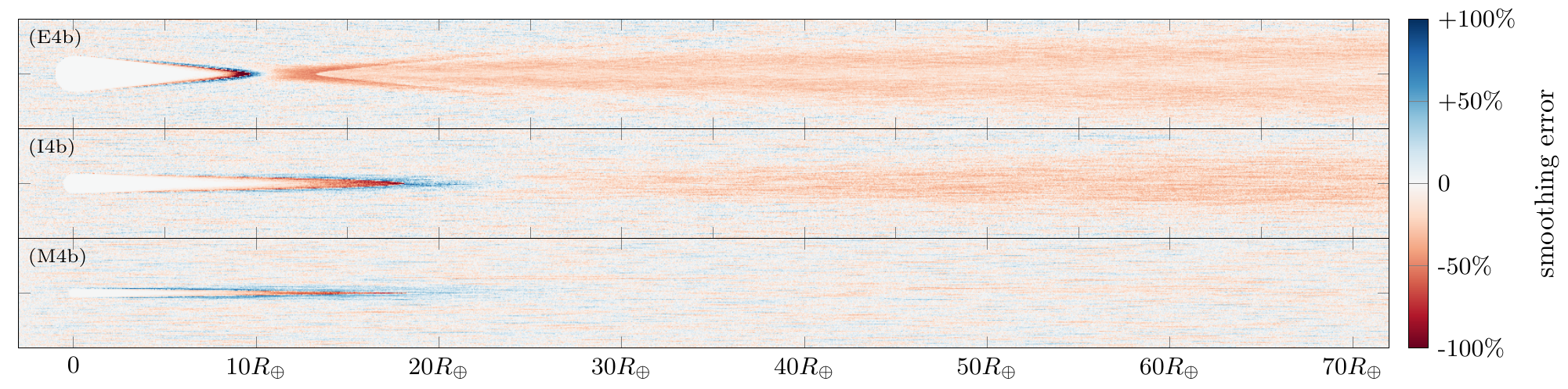}
    \caption{Percent difference between simulations and our analytic approximations (see appendix~\ref{apx:alg}).}
    \label{fig:diff}
\end{figure*}

However, given that the radiant dispersion produces a fuzzy boundary, it is not likely to be of much value to determine whether the observer is in the small region of space that lies outside the dispersion triangle but within the nominally shaded region. We therefore suggest treating this case as equivalent to the focusing-dominated case.

\subsection{Success of approximations}

In this section, we compare our approximations (equations~(\ref{eq:etagp}) and, where applicable, (\ref{eq:ylim})) with our simulations of meteoroid streams with non-zero radiant and speed dispersion. We perform these tests on simulations E4b and M4b from Fig.~\ref{fig:gem}, using the mode of the simulated radiant distribution, 0.8$^\circ$, as $\xi_\mathrm{min}$. In order to probe the ``intermediate'' case (section~\ref{sec:intermediate}), we include an additional massive body with no real-world analog whose mass is $0.11 M_\oplus$ and whose radius is $0.57 R_\oplus$. Fig.~\ref{fig:diff} presents the difference between our numerical simulations and our analytic approximations, expressed as a percentage. Table~\ref{tab:sims} lists the key differences between our various simulations.

\begin{table}
    \centering
    \begin{tabular}{cccc}
        simulation & central body & dispersion & $w$ (km~s$^{-1}$)\\
        \hline
        E1a & Earth & none & 25 \\ 
        E2a & Earth & radiant & 25 \\ 
        E3a & Earth & speed & 25 \\ 
        E4a & Earth & velocity & 25 \\ 
        E1b & Earth & none & 35 \\ 
        E4b & Earth & velocity & 35 \\ 
        I4b & intermediate & velocity & 35 \\ 
        M1b & Moon & none & 35 \\ 
        M4b & Moon & velocity & 35 
    \end{tabular}
    \caption{Summary of simulations presented in this paper. Note that ``intermediate'' refers to a central body with no real-life analog whose mass is 0.11 $M_\oplus$ and radius is 0.57 $R_\oplus$. A ``velocity'' dispersion indicates that each vectorial component of $\vec{w}$ follows a Gaussian distribution with $\sigma = 0.5$~km~s$^{-1}$, while a ``speed'' dispersion indicates that only the magnitude, $w$, is allowed to vary.}
    \label{tab:sims}
\end{table}

When $\xi$ is large (i.e., when the observer lies far from the anti-radiant line), the simulation and equation agree very well; the only visible differences are due to noise in the numerical simulation. For smaller values of $\xi$, our equation tends to underestimate the gravitational enhancement in the focusing-dominated and intermediate cases. In the region where focusing is the most intense, we underestimate the degree of focusing by about 25\%. We have used the mode of the Rayleigh distribution -- 0.8$^\circ$ -- as our value of $\xi_\mathrm{min}$. We tested other values, such as the square root of the variance of the distribution, but none performed better than the mode. Thus, an approximation that uses only the mode of the (Rayleigh) radiant dispersion is able to describe the pattern of gravitational focusing produced by both this Rayleigh radiant dispersion and a corresponding speed dispersion.

In the regions where one or both trajectories are blocked according to \S\ref{sec:shielding}, we underestimate the number density by up to 100\%. This is because meteoroids can ``bleed'' into these regions in our simulations in a way that our analytic approximations do not capture. However, the physical extent of these regions is small and limited to the border of the shadow.

Overall, we deem our approximations to be in good agreement with our simulations: a 25\% error in the number density enhancement is a significant improvement over the many-orders-of-magnitude errors that can occur when neglecting the radiant dispersion entirely. 

\subsubsection{The \"{O}pik test}

In this paper, we are primarily concerned with reproducing the local gravitational focusing at a single observer's location. However, if one instead considers the total flux incident on the surface of a planet, a much simpler equation applies. \cite{1951PRIA...54..165O} showed that, for a given value of $w$, gravity increases the flux incident on a planet's surface by a factor of:
\begin{align}
    \frac{
        \oint{|w u_r| \cdot H(-u_r) \cdot \eta(\xi)}
    }{
        \oint{|w_r| \cdot H(-w_r)}
    } &= 1 + 2f
    \label{eq:opik}
\end{align}
where $H$ is the Heaviside function and $f$ at the planet's surface is of course $G M/R w^2$. Equation~(\ref{eq:opik}) can thus be used to test whether a set of gravitational focusing equations behave as expected.

\cite{2007MNRAS.375..925J} were not able to satisfy the ``\"{O}pik test'' using equation~(\ref{eq:etag}); see Fig.~6 of \cite{2007MNRAS.375..925J}. This appears to be due to a misunderstanding; equation~(\ref{eq:etag}) describes the factor by which the \emph{number density} of meteoroids is increased due to gravitational focusing. It is not equivalent to the \emph{flux} enhancement factor described by \cite{1951PRIA...54..165O}; the flux is further enhanced due to the local increase in meteoroid speed. \cite{Moorhead2019b} demonstrate that equation~(\ref{eq:etag}) does, in fact, satisfy the \"{O}pik test when applied as a number density enhancement factor.

\begin{figure}
    \centering
    \includegraphics[width=3.31in]{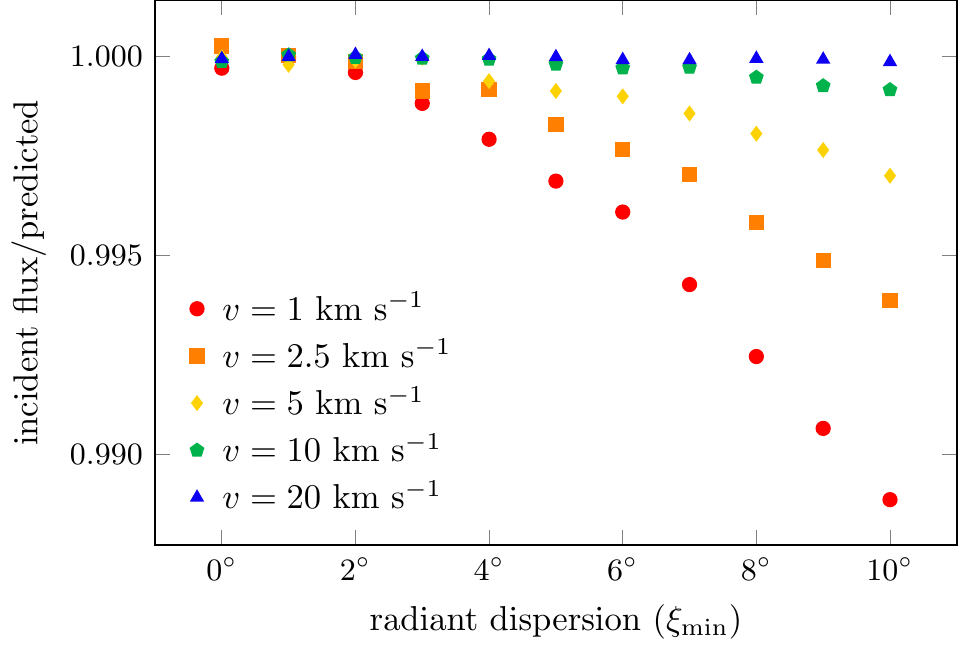}
    \caption{The flux incident on the top of the Earth's atmosphere, expressed as a fraction of the expected values (see equation~\ref{eq:opik}. This fraction is presented as a function of radiant dispersion angle for several choices of initial meteoroid velocity.}
    \label{fig:opik}
\end{figure}

We do not expect satisfaction of the \"{O}pik test to be much compromised by our smoothing terms. The largest modification to the number density or flux enhancement occurs where $\xi$ is small, and these locations will not experience any incoming flux due to planetary shielding. Nevertheless, as a test, we compute the total flux incident on the top of the Earth's atmosphere (i.e., we compute the sum of $- \eta' u_r w$ where $r = R_\oplus$ and $u_r < 0$) for a variety of initial meteoroid speeds and dispersion angles and compare it with the expected value. Except where the initial speed is very low and the dispersion very high, the results satisfy the \"{O}pik test to within 1\% (see Fig.~\ref{fig:opik}).

\subsubsection{Conservation of flux}

Perhaps a better test of the ramifications of our modified gravitational focusing treatment is whether the total flux in the absence of planetary shielding is zero. 

\begin{figure}
    \centering
    \includegraphics[width=3.31in]{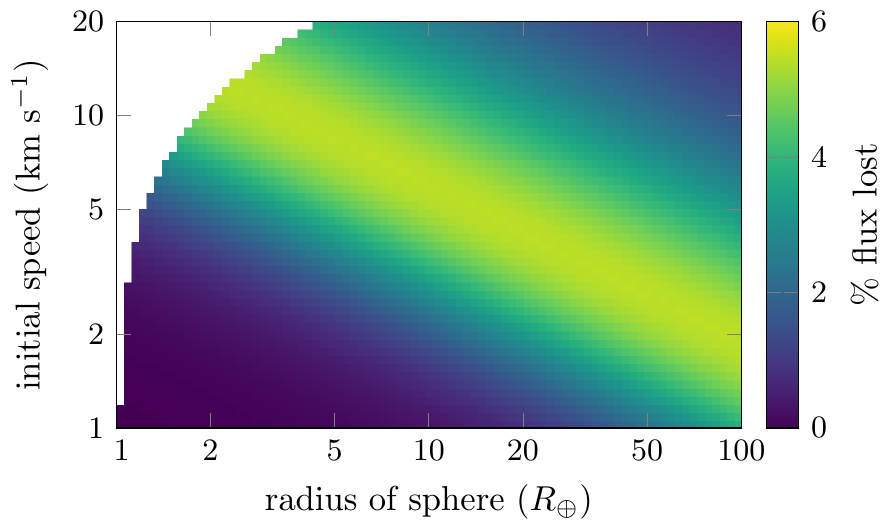}
    \caption{Imbalance between the outgoing flux and incoming flux on a sphere surrounding the Earth, neglecting planetary shielding and assuming a radiant dispersion of 5$^\circ$.}
    \label{fig:fluxcon}
\end{figure}

Fig.~\ref{fig:fluxcon} displays the results of a series of such tests. In these tests, we place a sphere around the Earth and compute the total flux on the surface of this sphere:
\begin{align}
    \oint{|w u_r| \cdot \eta'(\xi)}
    \label{eq:ftot}
\end{align}
If we were to integrate $\eta$ rather than $\eta'$, we would obtain a total flux of zero. However, since the number density of material passing out of the sphere at low $\xi$ tends to be artificially reduced by our smoothing factor, we expect our total to be slightly negative. To put the flux imbalance into perspective, we divide by the total incoming flux and express this fraction as a percentage. We perform these calculations only for distances lying outside the planet's shadow; inside the shadow, low-$\xi$ locations are blocked and this test is not relevant.

\begin{figure}
    \centering
    \includegraphics[width=3.31in]{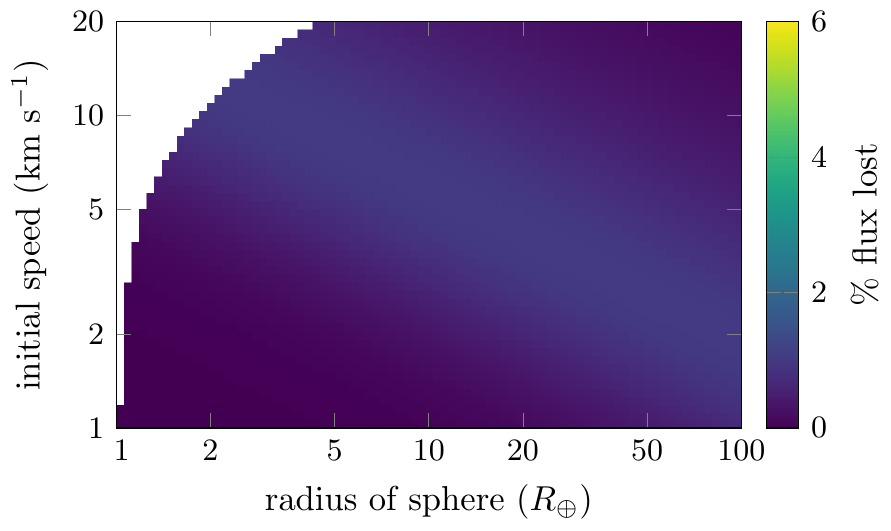}
    \caption{Imbalance between the outgoing flux and incoming flux on a sphere surrounding the Earth, neglecting planetary shielding and assuming a radiant dispersion of 1$^\circ$.}
    \label{fig:fcon1}
\end{figure}

Fig.~\ref{fig:fluxcon} shows that when the the incoming speed is low, up to 5\% of the flux can be lost at certain distances. However, for typical shower speeds -- those greater than 20~km~s$^{-1}$ -- the discrepancy is smaller. We have assumed a radiant dispersion of 5$^\circ$ in these simulations in order to exaggerate this effect; for a more typical dispersion of 1$^\circ$, the disagreement is at most 1\% (see Fig.~\ref{fig:fcon1}). Thus, at least in the case of the Earth, our analytic approximations are unlikely to introduce large errors in either the local number density or flux, the total flux incident on the Earth, or the total flux moving through a volume.

\subsubsection{Apparent radiant}

\begin{figure*}
    \centering
    \includegraphics{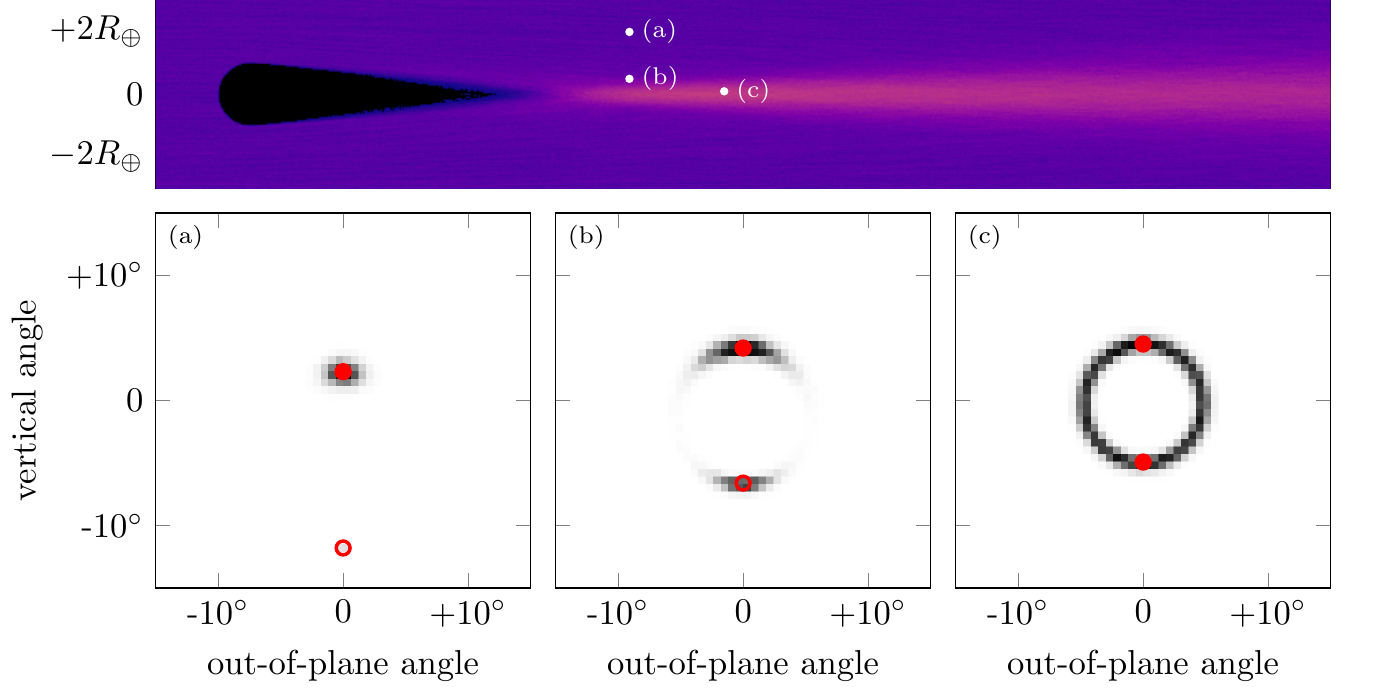}
    \caption{The apparent angular distribution with which meteoroids intersect an observer's location in simulation E4b. Observer locations relative to the Earth-like gravitating body are labeled in the top panel. The lower panels show the apparent radiant distribution seen when facing the radiant, where the vertical axis measures the vertical angle and the horizontal axis measures the angle away from the plane shown in the top panel. The grayscale heat map shows the radiant distribution produced by our simulation, and the red dots mark the nominal apparent radiants predicted by our analytic approximations. Open red dots indicate that the apparent radiant is predicted to be blocked by the massive body.}
    \label{fig:radplus}
\end{figure*}

Our modified version of the \cite{1997AdSpR..19..301S} approach still has only two paths by which meteoroids may reach the observer. However, a diffuse radiant permits an observer to ``see'' meteoroids coming from a ring or arc of apparent radiants. Figure~\ref{fig:radplus} shows the distribution of apparent radiants as seen at three observer locations in simulation E4b. Locations (a) and (b) lie in the region in which only one nominal trajectory is predicted to reach the observer, while location (c) lies very close to the anti-radiant line and is predicted to encounter meteoroids from both the ``long'' and ``short'' paths.

When the observer lies far from the anti-radiant, the ``short'' path is a reasonable description of the apparent meteoroid directionality. The vast majority of apparent radiants lie close to the predicted encounter angle. A few meteoroids appear to come from the ``long'' path, but this is a small fraction of the total. The agreement worsens as we approach the anti-radiant. Location (b) lies near the border of the region in which both trajectories are not blocked, but the long path is still predicted to be blocked. As a result, a substantial fraction of the meteoroids appear to lie near the long path. At locations even closer to the anti-radiant, such as (c), both trajectories are predicted to contribute, but the apparent radiant distribution resembles a ring rather than two points or arcs.

Thus, our approach is not capable of describing the extended directionality of meteoroids with non-zero radiant dispersion intersecting observer locations near the anti-radiant line. A full description of the directionality can likely only be obtained by a more computationally intensive approach that considers the full distribution of meteoroid orbits.

\section{Applications}

In this section, we discuss the ramifications of our study for several different scenarios.

\subsection{Meteoroid environment models}

We anticipate that the primary application of our numerical approximations will be in environment models. 
For instance, the NASA Meteoroid Environment Office (MEO) issues meteor shower forecasts \citep{2019JSpRo..56.1531M} that have been recently updated to incorporate the \cite{1997AdSpR..19..301S} treatment of gravitational focusing. We have found that spacecraft at high altitudes, such as those in geostationary orbit, often pass near the anti-radiant of active meteor showers and were thus predicted to encounter brief but intense spikes in the impactor flux. In this paper, we have shown that these spikes are not realistic, given our best estimates of meteor shower radiant dispersions. The modified algorithms we have derived here will be incorporated into future spacecraft-specific forecasts.

Our algorithms may also be useful for sporadic meteoroid models. For example, the MEO also generates a piece of software called the Meteoroid Engineering Model \citep[MEM][]{Moorhead2019b} that describes the sporadic meteoroid flux encountered along a spacecraft trajectory. MEM models the environment as a collection of meteoroid orbits that contribute to the local number density. If the velocity corresponding to one of these orbits is very closely parallel to the separation between the Earth and spacecraft, the anti-radiant singularity could produce what appears to be a ``hot pixel'' in the resulting flux map. Assigning some angular width to the apparent radiant would eliminate this possibility.

Neither MEM nor the MEO's shower forecasts model meteoroids whose speed relative to the massive body at infinity is less than 1~km~s$^{-1}$. However, models of asteroidal material or dust particles may include particles on highly circularized orbits that encounter the Earth at extremely slow speeds. Several such studies \citep{1985Icar...63..290W,2013Icar..226.1550K,2019JGRE..124..752P} have noted that gravitational focusing enhancement of these particles can be extreme. For instance, \cite{2013Icar..226.1550K} noted that low-inclination, quasi-satellite dust particles have enhancement factors of 3000. To avoid this, \cite{2019JGRE..124..752P} applied a softening parameter of 0.1~km~s$^{-1}$ to \"{O}pik's relation. The minimum speed we derived from the Hill radius -- 0.048~km~s$^{-1}$ -- is only a factor of two smaller and thus supports the decision made by \cite{2019JGRE..124..752P} to smooth over this velocity scale.

\subsection{Gravitational enhancement of showers by the Earth or Moon}

As discussed in section~\ref{sec:sims}, it has been hypothesized that gravitational focusing by the Moon could produce enhanced meteor shower activity at the Earth (M.\ Matney, personal communication) or by the Earth at the Moon (see Fig.~\ref{fig:gem}). We discussed the case of the Geminids, but, to give another example, in some years the Moon passes within a few degrees of the kappa and lambda Virginid (KVI and LVI) anti-radiants. In the absence of any radiant dispersion, such an alignment would result in brief, intense enhancements of these meteor showers.

Fig.~\ref{fig:gem}(M4b) reveals that the Moon is highly unlikely to produce any observable modification of meteor rates at Earth. However, Fig.~\ref{fig:gem}(E4b) indicates that gravitational focusing of a stream by the Earth could enhance the rate of meteoroids striking the Moon by a factor of a few. The meteoroid flux at the Moon can be measured via impact flashes \citep{2014Icar..238...23S,2014MNRAS.439.2364M,2016SPIE.9911E..22B}; determining whether and when such enhancements are visible could be worthy of further study.

\subsection{Second-order focusing in the Earth-Moon system}

Any meteoroid within the Moon's Hill sphere is also necessarily within the Earth's Hill sphere. To calculate the trajectory of such a particle with absolute accuracy, it is therefore necessary to compute the effects of both the Earth and the Moon. Similarly, a stream of meteoroids will experience second-order gravitational focusing in which the flux and number density are modified by both massive bodies. 

However, given the manner in which a modest radiant dispersion reduces the gravitational enhancement produced by the Earth at the Moon's location, and given also that scattering by the Earth will tend to increase dispersion in the meteoroid stream, we conclude that the effects of second-order focusing can be ignored.

\section{Conclusions}

While the effect of a massive body on a meteoroid stream can be described analytically, these equations contains a singularity that results in an arbitrarily large enhancements along the anti-radiant line. An additional singularity causes similarly large enhancements as the initial speed of the meteoroids approaches zero. A real meteoroid stream, however, will exhibit a dispersion in its direction of motion that eliminates these singularities. 

We found that meteoroid radiant dispersions can be described as a Rayleigh distribution and that the mode of this distribution can thus be used to describe the level of dispersion. We measured the dispersion of two showers and found that the Orionids had a radiant dispersion of 0.5$^\circ$ and the Perseids 1.2$^\circ$. In the absence of measurements, we suggest assuming radiant dispersions of roughly 0.5-1$^\circ$. Additional radiant dispersion measurements are the subject of a future paper. Our approach assumes that the dispersion in the velocity vector of the stream is isotropic. If the dispersion in the transverse component of the meteoroids' speeds is larger or smaller than the dispersion in the speed in the direction of the stream's motion, or if the radiant distribution is not symmetric, this assumption is not valid.

We simulated the effect of modest radiant (and speed) dispersions on the behavior of a stream of meteoroids passing near a massive body. We assumed a 0.5~km~s$^{-1}$ dispersion in speed, corresponding to a Rayleigh radiant dispersion with a mode of 1.15$^\circ$ for an average speed of 25~km~s$^{-1}$ or 0.8$^\circ$ for 35~km~s$^{-1}$. In all cases, we found that including a dispersion in radiant substantially reduced the maximum local number density enhancement. We also found that it limited the length of the region shielded from the stream by the massive body.

We derived modifications of the \cite{1997AdSpR..19..301S} treatment of gravitational focusing and planetary shielding that approximate the effects of a radiant dispersion. These approximations, which are summarized in appendix~\ref{apx:alg}, agree with our simulations to within 25\%. We found that the mode of the Rayleigh radiant distribution worked well as a smoothing factor, which we call $\xi_\mathrm{min}$. An approximation using only the mode of the radiant distribution was able to describe the pattern of gravitational focusing produced by meteoroids with dispersions in both their speeds and radiants. Tests of our approximations showed that our modified gravitational focusing equations satisfy the so-called \"{O}pik test to within 1-2\%, and conserve flux to within a few percent.

We anticipate that our algorithms could be useful in meteoroid environment modeling, where an efficient yet accurate description of the flux near a massive body is desired. Shower models would benefit from individual measurements of radiant dispersion; we plan to conduct a survey of meteor shower radiant dispersions in the near future.





\section*{Acknowledgements}

This work was supported in part by NASA Cooperative Agreement 80NSSC18M0046, by the Natural Sciences and Engineering Research Council of Canada, and by Jacobs contract 80MSFC18C0011. Initial funding for the 16~mm GMN stations was provided by the Istria County Association of Technical Culture.

The authors would like to thank the following GMN station operators and contributors whose stations provided the data used in this work (in alphabetical order): \v{Z}eljko Andrei\'{c}, Ricky Bassom, Richard Bassom, Jean-Philippe Barrilliot, Josip Belas, Martin Breukers, Seppe Canonaco, Jose Carballada, Dino \v{C}aljku\v{s}i\'{c}, Ivica \'{C}ikovi\'{c}, J\"{u}rgen D\"{o}rr, Alfredo Dal\'{}Ava J\'{u}nior, Jean-Paul Dumoulin, Peter Eschman, Jim Fordice, Jason Gill, Nikola Gotovac, Tioga Gulon, Margareta Gumilar, Pete Gural, Bob Hufnagel, Jean-Marie Jacquart, Ron James, Vladimir Jovanovi\'{c}, Milan Kalina, Steve Kaufman, Zoran Knez, Dan Klinglesmith, Danko Ko\v{c}i\v{s}, Korado Korlevi\'{c}, Stanislav Korotkiy, Zbigniew Krzeminski, Mirjana Malari\'{c}, Nedeljko Mandi\'{c}, Michael Mazur, Aleksandar Merlak, Matej Mihel\v{c}i\'{c}, Gene Mroz, Przemek Naga\'{n}ski, Jean-Louis Naudin, Damjan Nemarnik, Zoran Novak, Thiago Paes, Lovro Pavleti\'{c}, Enrico Pettarin, Alan Pevec, Danijel Reponj, Paul Roggemans, James Rowe, Dmitrii Rychkov, Jim Seargeant, Bela Szomi Kralj, Damir \v{S}egon, Rajko \v{S}u\v{s}anj, Yakov Tchenak, William Wallace, Steve Welch, Alexander Wiedekind-Klein, and Dario Zubovi\'{c}.  The authors also thank W.\ J.\ Cooke and P.\ G.\ Brown for their support and guidance, and Auriane Egal for helpful discussions about physical radiant dispersions of meteor showers.

Finally, the authors thank an anonymous reviewer for independently reproducing and checking our modeling results and for prompting us to include sections 4.3.3 and appendix A.


\bibliographystyle{mnras}
\bibliography{local} 

\appendix

\section{Meteor velocity and radiant probability distributions}
\label{apx:dists}

In this paper, we assume that the distribution of the Cartesian components of meteoroid velocities within a shower can each be described by a normal distribution:
\begin{align}
    p_x (v_x) &= \frac{1}{\sigma \sqrt{2 \pi}}
        e^{-v_x^2/2 \sigma^2} \\
    p_y (v_y) &= \frac{1}{\sigma \sqrt{2 \pi}}
        e^{-v_y^2/2 \sigma^2} \\
    p_z (v_z) &= \frac{1}{\sigma \sqrt{2 \pi}}
        e^{-(v_z - v_0)^2/2 \sigma^2} 
\end{align}
where $p_{v_i}$ gives the probability of selecting speed $v_i$. Here, we have defined the $z$ direction to point in the direction of the nominal shower velocity.

The ratio of the square of the magnitude of the corresponding velocity vector,
\begin{align}
    (v/v_0)^2 &= (v_x^2 + v_y^2 + v_z^2)/v_0^2 \, ,
\end{align}
then follows a noncentral chi-squared distribution with three degrees of freedom and a noncentrality parameter of $\lambda = (v_0/\sigma)^2$. When $\lambda$ is large -- that is, when the velocity dispersion within a stream is small compared to the overall speed -- the probability distribution of $v$ can be approximated as a normal distribution:
\begin{align}
    p_v (v) &\simeq \frac{1}{\sigma \sqrt{2 \pi}}
        e^{-(v - v_0)^2/2 \sigma^2} 
\end{align}
One such approximate distribution is shown in Fig.~\ref{fig:vhist}.

The component of the speed that is transverse to the motion of the stream,
\begin{align}
    v_\bot &= \sqrt{v_x^2 + v_y^2} \, ,
\end{align}
follows a Rayleigh distribution with a mode equal to $\sigma$:
\begin{align}
    p_\bot(v_\bot) &= \frac{v_\bot}{\sigma^2}
        e^{-v_\bot^2/2 \sigma^2} \, . \label{eq:pbot}
\end{align}
The radiant offset is then
\begin{align}
    \alpha &= \tan^{-1} \frac{v_\bot}{v_z} \, .
\end{align}
If $v_0 \gg \sigma$, then $v_z \simeq v_0$, $v_\bot \ll v_z$, $\tan \alpha \ll 1$, and $\alpha \simeq \tan \alpha$. In such a case, the radiant offset then also follows a Rayleigh distribution:
\begin{align}
    p_\alpha(\alpha) &= \frac{\alpha}{\sigma_\alpha^2}
        e^{-\alpha^2/2 \sigma_\alpha^2} \, , \label{eq:palpha}
\end{align}
where the mode is $\sigma_\alpha \simeq \sigma/v_0$.

A small dispersion in the transverse velocity is necessary for a group of meteors to be recognized as a shower. We will therefore assume that equations~(\ref{eq:pbot}) and (\ref{eq:palpha}) apply throughout this paper.


\section{Using the Hill radius to place a limit on gravitational focusing}
\label{apx:hill}

In this section, we derive a speed smoothing factor by assuming that meteoroids outside the massive body's Hill radius do not contribute to gravitational focusing inside the Hill sphere.

The impact parameter of a particle intersecting the observer with angle $\phi$ is \citep{2007MNRAS.375..925J}:
\begin{align}
    x &= r \sin \phi \sqrt{ 1 + 2 f }
\end{align}
Note that $\phi$ describes the direction of the particle's motion \emph{after} gravitational focusing, while $\xi$ describes the direction of motion prior to focusing. Note also that we use $x$ to denote impact parameter -- the close-approach distance between the meteoroid and the massive body in the absence of gravity -- rather than the more commonly used symbol $b$.

For a negligible initial speed, the maximum impact parameter of any particle intersecting the observer's location ($x_\mathrm{max}$) corresponds to $\sin \phi = 1$:
\begin{align}
    x_{\mathrm{max}} &= r \sqrt{ 1 + 2 f } 
\end{align}

If we limit the flux to those meteoroids passing through the Earth's Hill sphere ($x_{\mathrm{max}} = r_H = 0.01$~au), and assume that $r \ll r_H$, we obtain:
\begin{align}
    w_{\mathrm{min}}^2 &\simeq 2 G M r / r_H^2
\end{align}
At an altitude of 100~km above the Earth's surface, where the atmosphere is thick enough to ablate meteoroids, $r = R = 6471$~km and $w_{\mathrm{min}}$ is 48~m~s$^{-1}$. 

\section{An efficient gravitational focusing and shielding algorithm}
\label{apx:alg}

This appendix consolidates our analytic equations for determining the meteoroid number density enhancement due to gravitational focusing and planetary shielding. They are presented in an order and format that facilitates their translation to a computer programming language. 

We assume that the user knows: the gravitational constant, $G$; the mass of the massive body, $M$; the meteoroid-blocking radius of the massive body, $R$; a vector that describes the motion of the meteoroids prior to entering the massive body's sphere of influence, $\vec{w}$; a vector that describes the position of the observer relative to the center of the massive body, $\vec{r}$; a vector that describes the velocity of the observer relative to the massive body, $\vec{v}$; and the mode of a Rayleigh distribution describing the meteoroid stream's radiant dispersion, $\xi_\mathrm{min}$. This latter variable is not routinely measured for meteor showers and the user may therefore need to obtain a new measurement of the radiant dispersion or estimate it from similar showers.

\vspace{0.1in}
\noindent
Step 1. Convert the radiant dispersion angle to a smoothing scale:
\begin{align}
    s' &= \sin^2 \tfrac{\xi_\mathrm{min}}{2}
\end{align}

\vspace{0.1in}
\noindent
Step 2. Determine the magnitude of the initial meteoroid speed:
\begin{align}
    w &= \lVert \vec{w} \rVert
\end{align}

\vspace{0.1in}
\noindent
Step 3. Determine the distance of the observer from the center of the massive body:
\begin{align}
    r &= \lVert \vec{r} \rVert
\end{align}

\vspace{0.1in}
\noindent
Step 4. Calculate the semi-major axis of the meteoroid stream:
\begin{align}
    a &= - G M/w^2
\end{align}

\vspace{0.1in}
\noindent
Step 5. Determine whether the scenario is shielding-dominated:
\begin{align}
    \sin \xi_\mathrm{min} &> \frac{2 a}{2 a - R}
\end{align}

\vspace{0.1in}
\noindent
Step 6. Compute the ratio of the massive body's meteoroid-blocking radius to $r$:
\begin{align}
    p &= R/r
\end{align}

\vspace{0.1in}
\noindent
Step 7. Find the cosine of the angle between the initial meteoroid velocity vector and the observer's geocentric position:
\begin{align}
    \cos \xi &= (\vec{w} \cdot \vec{r})/(w r)
\end{align}

\vspace{0.1in}
\noindent
Step 8. Compute the two half-angles:
\begin{align}
    s &= \sin^2 \tfrac{\xi}{2} = (1 - \cos{\xi})/2 \\
    c &= \cos^2 \tfrac{\xi}{2} = 1-s
\end{align}

\vspace{0.1in}
\noindent
Step 9. Compute the dimensionless term $f$:
\begin{align}
    f &= -a / r
\end{align}

\vspace{0.1in}
\noindent
Step 10. Compute the unitless ``auxiliary speed:''
\begin{align}
    b &= \sqrt{s (s + 2 f)}
\end{align}

\vspace{0.1in}
\noindent
The following steps must be completed for both the ``long'' and ``short'' paths. In other words, evaluation the following equations for both $k = -1$ and $k = 1$.

\vspace{0.1in}
\noindent
Step 11. Compute the radial component of the local meteoroid velocity, scaled to the initial speed $w$:
\begin{align}
    u_r &= c - k \cdot b
\end{align}

\vspace{0.1in}
\noindent
Step 12. Determine the magnitude of the tangential component of the local meteoroid velocity, scaled to the initial speed $w$:
\begin{align}
    u_\bot &= 1 + 2 f - u_r^2
\end{align}

\vspace{0.1in}
\noindent
Step 13a. If the scenario is not shielding-dominated, determine whether the meteoroid trajectory in question is blocked by the massive body as follows:
\begin{align}
    u_r &> 0 \\
    u_\bot^2 &< p^2 + 2 f p
\end{align}
If both of the above statements are true, the trajectory is blocked. (Note that if $r < R$, the trajectory is also blocked.)

\vspace{0.1in}
\noindent
Step 13b. If the scenario is shielding-dominated, determine whether the meteoroid trajectory in question is blocked by the massive body as follows:
\begin{align}
    \sin \xi + \cos \xi \tan \xi_\mathrm{min} < p
\end{align}
If the above statement is true, the trajectory is blocked. (Note that if $r < R$, the trajectory is also blocked.)

\vspace{0.1in}
\noindent
Step 14. If the trajectory is blocked, $\eta' = 0$. Otherwise, 
\begin{align}
    b' &= \sqrt{(s + s') (s + s' + 2 f)} \\
    \eta' &= \tfrac{1}{2} \left| \, 1 + k (s + s' + f) / b' \right|
\end{align}
Sum both possible values of $\eta'$ to obtain the total number density enhancement factor.

\vspace{0.1in}
\noindent
Step 15. If desired, compute the velocity of the meteoroid relative to the observer.
\begin{align}
    w \vec{u} - \vec{v} &= w u_r \hat{r} + w \sqrt{u_\bot^2} \hat{u}_\bot - \vec{v} \\
    \hat{u}_\bot &=  
    \frac{\vec{r} \times (\vec{r} \times \vec{w})}{r^2 w \sqrt{1 - \cos^2 \xi}}
\end{align}
Note that the two paths will produce two separate apparent radiants at the observer's location.

\vspace{0.1in}
\noindent
Step 16. If desired, compute the meteoroid flux enhancement factor:
\begin{align}
    \zeta' &= \eta' \frac{\lVert w \vec{u} - \vec{v} \rVert}{\lVert \vec{w} - \vec{v} \rVert}
\end{align}
As with the number density enhancement factor, sum the two possible values of $\zeta$ to obtain the total flux enhancement factor.



\bsp	
\label{lastpage}
\end{document}